\documentclass[twocolumn,showpacs,prl,superscriptaddress]{revtex4-1}

\usepackage{amsmath}
\usepackage{tabularx}
\usepackage{graphicx}
\usepackage{color}
\usepackage{times}

\graphicspath{{data_phil/}{data/}}

\begin{document}

\title{Replica symmetry breaking for Ulam's problem}

\author{Phil Krabbe}
\affiliation{Institut f\"ur Physik, Universit\"at Oldenburg, D-26111
	     Oldenburg, Germany}
\author{Hendrik Schawe}
\affiliation{Laboratoire de Physique Th\'{e}orique et Mod\'{e}lisation,
UMR-8089 CNRS-Universit\'{e} Cergy Pontoise, France}
\author{Alexander K. Hartmann}
\affiliation{Institut f\"ur Physik, Universit\"at Oldenburg, D-26111
	     Oldenburg, Germany}

\begin{abstract}
We study increasing subsequences (IS)
for an ensemble of sequences given by
permutation of numbers $\{1,2,\ldots,n\}$. We consider a Boltzmann
ensemble at temperature $T$.
Thus each IS appears with the corresponding Boltzmann
probability where the  energy is the negative length $-l$ of the IS. 
For $T\to 0$,
only ground states, i.e. longest IS (LIS) contribute,
also called Ulam's problem. We introduce
an algorithm which allows us to directly sample IS in perfect equilibrium
in polynomial time, for any given sequence and any temperature. Thus,
we can study very large sizes. We obtain averages for the first and second
moments of number
of IS as function of $n$ and confirm analytical predictions.
Furthermore, we analyze for low temperature $T$
the  sampled ISs by computing
the distribution of overlaps and performing hierarchical cluster analyses.
In the thermodynamic limit $n\to \infty$ the distribution of overlaps
stays broad and the configuration landscape remains complex. Thus, Ulam's
problem  exhibits replica symmetry breaking. This means it
constitutes a model with complex behavior which can be
studied numerically exactly in a highly efficient way, in contrast to other
RSB-showing models, like spin glasses or NP-hard optimization problems,
where no fast exact algorithms are known.
\end{abstract}


\maketitle

{\it (Introduction) }
The mathematician Stanis\l{}av Ulam was also a well-known pioneer
in Computer simulations.
One of the problems he studied numerically, back in the 1950s,
 was \cite{beckenbach2013modern} the scaling of the length $L$
of the longest increasing subsequence (LIS)
\cite{romik2015} of random permutations of $n$  numbers.
Based on the knowledge \cite{erdoes1935} that the average length
increases at least like the square root of $n$,
he proposed that the average
length scales as $\langle L \rangle = c\sqrt{n}$ with $c\approx 1.7$.
In the meantime,
$\langle L \rangle =2\sqrt{n}$ for $n\to\infty$
has been proven \cite{aldous1999longest}.
 Also the
distribution $P(l)$ of maximum lengths has been studied analytically
\cite{Seppalainen1998,logan1977variational,deuschel1999increasing}
and it was found \cite{baik1999distribution} that the central part
is given by the Tracy-Widom distribution. This was confirmed
numerically by large-deviations simulations \cite{lis2019}, and
also considered for other sequence ensembles  \cite{studentlis2020}. Furthermore,
 the expectation values of the
 number of IS of a certain length \cite{hammersley1972} and of
 all increasing subsequences (IS) \cite{lifschitz1981} have been obtained
analytically. The actual distribution of the number of LIS was
obtained numerically over a large range of the support again by
applying large-deviation algorithms \cite{lis_count2020}.

As a tool, the calculation of LIS
finds also applications outside mathematics, like in data analysis
\cite{gopalan2007}, financial fraud detection \cite{bonomi2016differentially},
or sequence alignment in bioinformatics \cite{Zhang2003}.

In spite of these connections to many fields,
to our knowledge, the behavior of IS and LIS was studied
so far surprisingly
only with respect to the length  and to the exponentially growing
number  of increasing
subsequences. Thus, we are not aware of any study, where the actual structure
of the exponentially large IS configuration space has been studied.

Such questions with respect to phase-space organization
lie at the heart of the statistical mechanics of complex systems
like glasses, spin glasses, machine
learning or optimization problems
\cite{binder1986,mezard1987,fischer1991,young1998,nishimori2001,phase-transitions2005,mezard2009,moore2011,kawashima2013}. In particular one is
interested whether the configuration space is rather simple, like
for a ferromagnet, often coined as \emph{replica symmetric}, or
whether it is complex with a hierarchical organization of
phase space coined as \emph{replica symmetry breaking} (RSB), as it
appears for mean-field spin glasses \cite{parisi1979}.

In most cases, analytical solutions cannot be obtained, so one has
to use computer simulations \cite{practical_guide2015}.
Unfortunately,
 all standard models where one knows or suspects that
they exhibit a complex RSB-like behavior, like spin glasses, are numerically
very hard to treat. Hence, only rather limited system sizes could be
considered when performing equilibrium sampling,
even when using special parallel computers like JANUS \cite{janus}.
Note that for combinatorial optimization problems like the Satisfiability
problem, in some cases efficient algorithms exist
\cite{mezard2002}. But they allow only
 to find \emph{some} solution, i.e., this sampling is not controlled. Thus,
these algorithms  do
 not allow to sample the configuration space in equilibrium which is
necessary to study the configuration-space structure. For this purpose
one has to use Monte Carlo Markov-chain sampling, which requires
equilibration and is slow therefore.
Note,
for directed polymers in random media, where indeed a fast polynomial
sampling is possible \cite{huse1985,kardar1985,kardar1987}, recently a
broad distribution of overlaps
and a complex hierarchy of configurations was found \cite{directed_rsb2022},
but this was the case only for ensembles which exhibit special
correlations in the disorder, not the ``natural'' uncorrelated one.

Here we introduce an algorithm which allows one to count the number of IS
for any given length $l$ as well as sampling IS exactly for any given
distribution which depends only on the IS length $l$, in particular
for any given length the sampling is uniform. Both calculation of the
numbers and the sampling can be performed in polynomial time, which
allows us to treat large systems exactly. We study
the sequence ensemble of random permutations, which does not exhibit
correlations and is the classical and most-studied ensemble for IS and LIS.
Our results indicate that the structure of the
configuration space exhibits properties of replica-symmetry breaking, i.e.,
a broad distribution of overlaps and a hierarchical clustering
of configurations, even in the thermodynamic limit $n\to \infty$.

Next, we present all necessary definitions and introduce the algorithms.
Then we show our results and finish by a summary and discussion.


{\it (Definitions and Algorithms) }
Let $\sigma=(\sigma_1,\sigma_2,\ldots,\sigma_n)$ be a sequence of $n$
distinct numbers.
A subsequence $\lambda =(\sigma_{i_1},\sigma_{i_2},\ldots,\sigma_{i_l})$
of length $l=l(\lambda)$
fulfills $1\le i_1<i_2<\ldots<i_l\le n$ and is called \emph{increasing}
if $\sigma_{i_j} < \sigma_{i_{j+1}}$ for all $j=1,\ldots,l-1$. To calculate
the longest among all possible IS, the \emph{patience sort algorithm}
\cite{aldous1999longest} is a popular choice which runs in polynomial
time. Recently, an extension was proposed \cite{lis_count2020}, which
allows one to calculate the number of LIS. Here, we introduce a further
extension and variant of the algorithm,
which enables one to count all IS and sample
them efficiently and exactly for any desired probability distribution
which depends on the IS length l.

Let $H$ be a \emph{precedence matrix}, which encodes possible joint occurrences
of entries $\sigma_i$ and $\sigma_j$ in an IS $\lambda$, i.e.,
\begin{equation}
H_{ij}=\left\{
\begin{array}{rl}
1 & {\rm if}\,  i<j\, {\rm and}\, \sigma_i < \sigma_j \\
0 & {\rm else}.
\end{array}
\right.
\end{equation}
Note that the matrix $H$ can be efficiently stored as a graph with
neighbor lists. To set up $H$, we run an extended variant of patience sort, which
gives also the length $L$ of the LIS and allows us to restrict the number of
candidates $i,j$ which have to be checked whether one has to assign $H_{ij}=1$.
Still, this requires $O(n^2)$ steps.

To count IS and LIS, we denote by $\Psi^l_j$ the number of IS of length $l$
which end at position $j$. Clearly, each single entry of $\sigma$
represents an IS of length $l=1$,
i.e., we have $\Psi^1_i=1$ for $i=1,\ldots,n$. Now, IS of length $l>1$
can be constructed by selecting a final entry $\sigma_{j}$
precedent by an IS of length $l-1$ where all entries are smaller
than $\sigma_{j}$ and appear before position $j$. For the number
of IS this turns into
\begin{equation}
\Psi^{l}_j = \sum_{i<j} H_{ij}\Psi^{l-1}_i\quad {\rm for}\, l=2,\ldots,L
\end{equation}
which can be computed in a convenient way recursively, i.e.,
by \emph{dynamic programming} in $O(n^2L)$. The total number of IS
of length $l$ is given by $\Psi^l=\sum_j \Psi^l_j$, where we also
include the empty subsequence $\Psi^0=1$. The total
number of IS is given by $\Psi=\sum_{l=0}^L \Psi^l$.

To sample an IS for given length $l$,
one starts by sampling the final entry $j$
which appears with probability $\Psi^l_j/\Psi^l$. Next,
the preceding entry $i$ is sampled among all possible predecessors $i<j$,
i.e., where $H_{ij}=1$. Each possible entry $i$ is selected with
probability $\Psi^{l-1}_i/\Psi^{l}_j$. This is continued iteratively
for length $l-2,l-3$, etc, always given the just sampled entry, until length
0 is reached. This algorithm takes $O(n\,l)$ steps.

The sampling can be easily extended to include any probability which
depends on the length. Here we take a physical viewpoint by considering
$E=-l$ as energy within the canonical
ensemble at temperature $T$, i.e., by using
probabilities $\sim \exp(l/T)$. Thus, for
an IS $\lambda$ we have the probability  given by
\begin{equation}
p(\lambda) = \exp(l(\lambda)/T)/Z,\quad Z=\sum_l \Psi^l \exp(l/T)\,.
\label{eq:boltzmann}
\end{equation}
This includes in particular all LISs for $T\to 0$. Sampling an IS now
consist of first drawing a length $l$ according the probabilities
$\Psi^l\exp(l/T)/Z$,  and then uniformly sampling an IS of length $l$
as explained before. Note that this approach is exact and direct,
i.e., for each run of the algorithm an independently sampled configuration
is returned. Our approach runs in polynomial time,
such that we can treat rather large systems in perfect equilibrium.


{\it (Results)}
We performed simulations \cite{practical_guide2015}
for ensembles of permutations of $n$ numbers in the range
$n=128$ to $n=8192$. We studied for all sizes  10000 realizations
of the disorder, i.e., independent permutations.
For comparison, we also considered in some
cases the
ordered sequence $\sigma^o=(1,2,\ldots,n)$.

\begin{figure}
\includegraphics[width=0.99\columnwidth]{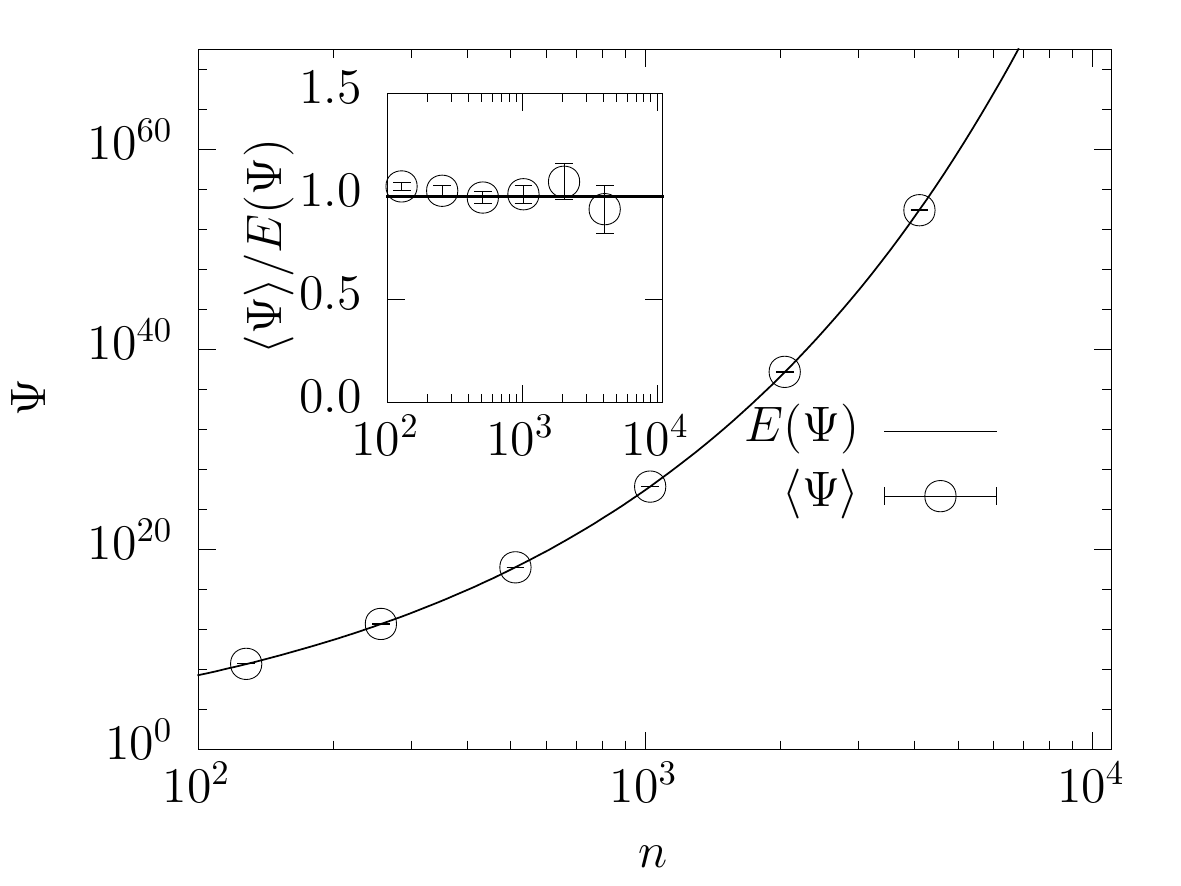}
\caption{\label{fig:num_is_N} The number $\Psi$ of IS as function of
sequence length $n$, for the analytical expectation value $E(\Psi)$
and  the numerical average $\langle \Psi \rangle$. The inset shows the
ratio $\langle \Psi \rangle / E(\Psi)$, the horizontal line is at value 1. }
\end{figure}

We start by considering the number $\Psi$ of IS. The asymptotic behavior
of the expectation value is analytically given by  \cite{lifschitz1981}
\begin{equation}
E(\Psi)= \frac{1}{2\sqrt{\pi e}} n^{-1/4}\exp(2\sqrt n).\
\end{equation}
In Fig.\ \ref{fig:num_is_N} we compare the numerical average
$\langle \Psi \rangle$ with the analytical result and find very good agreement,
even for rather small system sizes. Note that the average is ``annealed''
in the sense that is represents an exponentially growing quantity, such that
sequences with exceptionally large values of $\Psi$ will dominate.
This means, we need a rather large number of samples to observe
agreement, as we do. This also indicates the correctness of our approach.
 We have also evaluated
the second moment $\langle \Psi^2 \rangle$ (not shown).
 Here the agreement
with the analytical result \cite{lifschitz1981} is fair, i.e., a bit lower,
due to the even stronger
dominance of exponentially large but exponentially rare sequences. To find
a good agreement here, one would have to obtain the distribution $P(\Psi)$
down to the tails. This should be possible by using
 a large-deviation approach,
as it has been used to obtain the distribution of the number of LIS
\cite{lis_count2020},
but lies outside the scope of the present study.

Next, we analyze IS sampled in equilibrium according to
Eq.~(\ref{eq:boltzmann}) at a low temperature $T=0.2$
for several sequence lengths $n$. For independently sampled pairs
$\lambda^{(1)},  \lambda^{(2)}$ of IS, we calculate the similarity of the
two IS via the \emph{overlap} $q$.
Here, considering the IS as sets of the contained numbers, we use
the Jaccard-Index \cite{jaccard1912} as given by
\begin{equation}
q= \frac {|\lambda^{(1)} \cap  \lambda^{(2)}|}{|\lambda^{(1)} \cup  \lambda^{(2)}|}\,.
\end{equation}

\begin{figure}
\includegraphics[width=0.99\columnwidth]{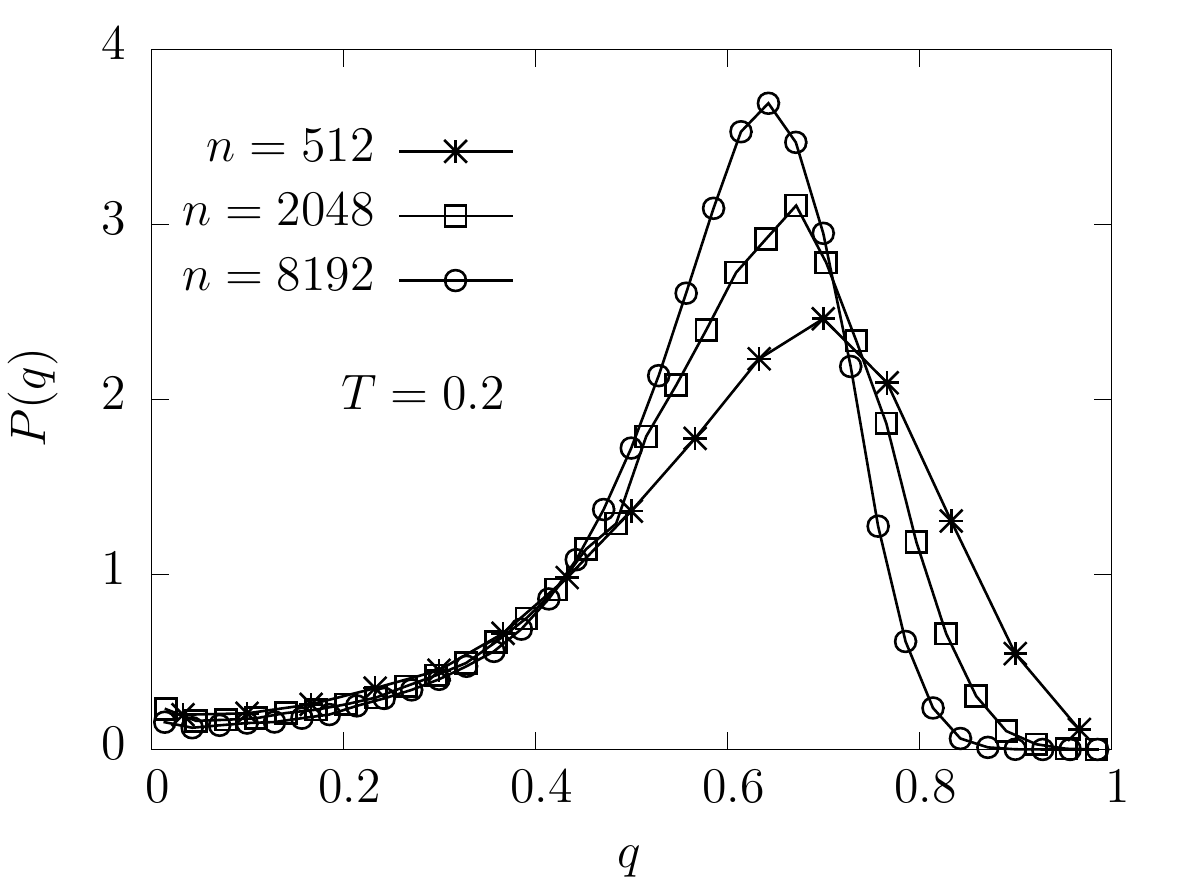}
\caption{\label{fig:distr_overlaps} Distribution of overlaps at $T=0.2$,
for three system sizes $n$. The lines are guides to the eyes only.}
\end{figure}

The distribution $P(q)$ of overlaps is shown for $T=0.2$
in Fig.~\ref{fig:distr_overlaps}
for three sequence sizes $n$. Apparently the distribution is broad, even
for large systems, indicating a complex configuration landscape.

\begin{figure}
\includegraphics[width=0.99\columnwidth]{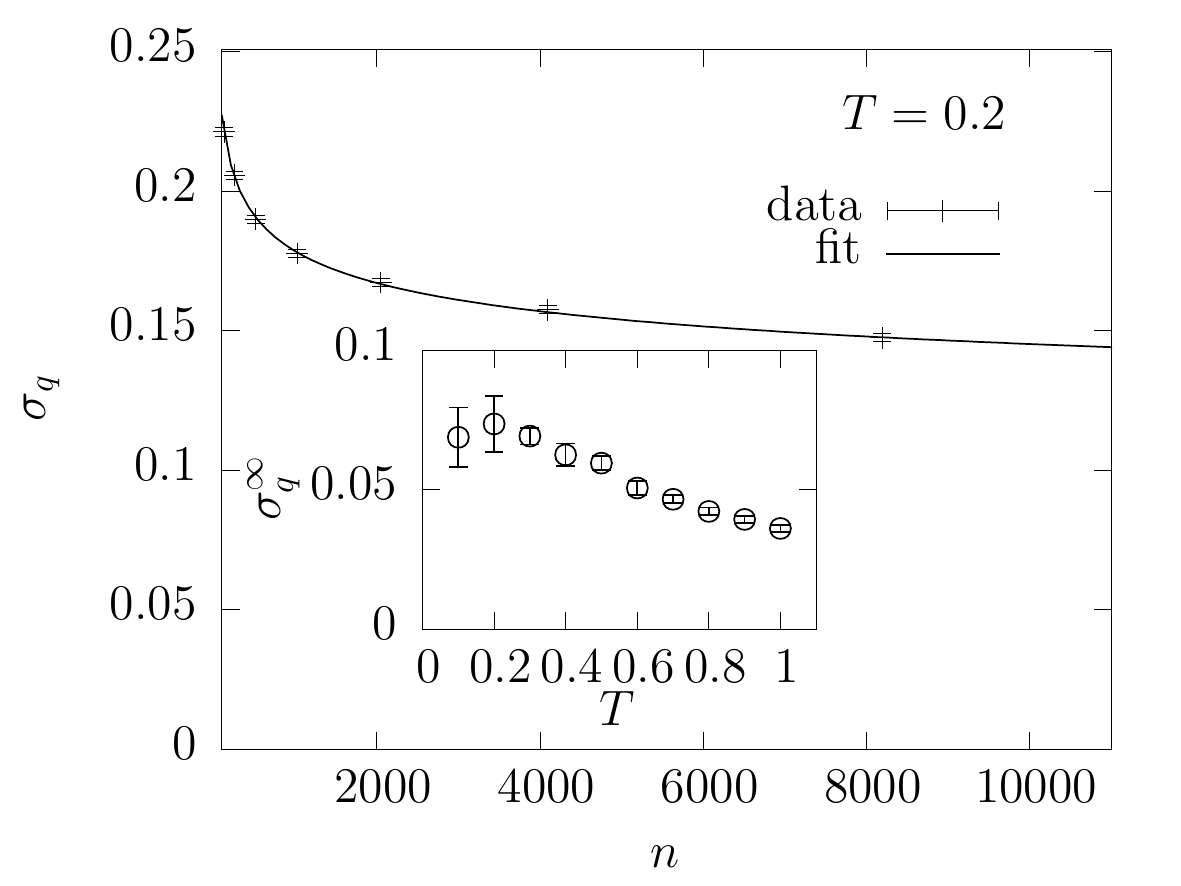}
\caption{\label{fig:sigma} Width $\sigma_q(n)$ of $P(q)$
at $T=0.2$. The line shows a fit of the data to a power law with offseet $\sigma_q(n)=\sigma_q^{\infty}+an^{-b}$.
The inset shows the extrapolated value
$\sigma_q^{\infty} = \lim_{n\to\infty} \sigma_q(n)$ as a function of the 
temperature.}
\end{figure}

To investigate whether this is true also in the thermodynamic limit,
we have evaluated the width $\sigma_q$ of the distribution as function
of system size $n$. The result for $T=0.2$ is shown in Fig.~\ref{fig:sigma}.
We fitted a power law $\sigma_q(n)=\sigma_q^{\infty}+an^{-b}$ and obtained
the limiting value $\sigma_q^{\infty}=0.073(10)$, which is significantly
different from zero and shows that the distribution remains broad
in the thermodynamic limit.

When evaluating $\sigma_q^{\infty}$ as function of $T$, see inset of
Fig.~\ref{fig:sigma}, it appears to be nonzero for all temperatures
in the studied range. Thus only a rather smooth
decrease is visible, no sign of a transition, where one would
expect a power-law decrease $\sim L^{-\eta}$ at and beyond the transition.
 This behavior we observe also
for the average overlap (not shown here).

\begin{figure}
\includegraphics[width=0.99\columnwidth]{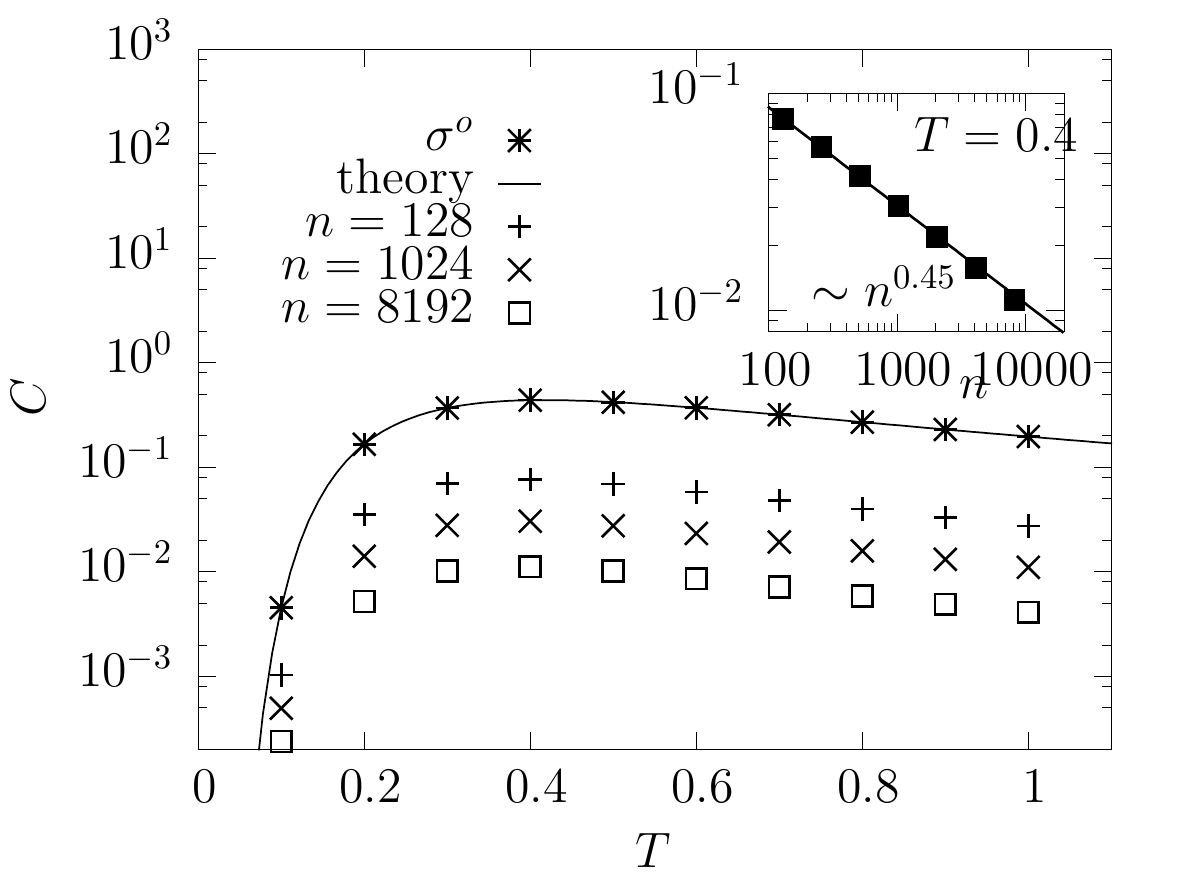}
\caption{\label{fig:specific_heat} The specific heat $C$ as function of
temperature $T$, for the ordered sequence $\sigma^0$, where
the line is the analytic result, and
averaged for the permutations. Shown are some system sizes $n$.
 The inset shows the scaling of the value near the peak for the permutations
as function of system size, following a power law $n^{-y}$, shown
as a straight line, with a fit value $y=0.451(5)$.}
\end{figure}

We also obtained
the specific heat via the variance of the length by calculating
$C= (\langle l^2 \rangle- \rangle l \rangle ^2)/(nT^2)$.
Interestingly, the disorder-averaged
$C(T)$ exhibits peaks near $T=0.4$ for all system sizes,
but the peak height decreases
with growing system size $n$. Thus, this behavior provides also no sign
for a phase transition.
Note that a non-growing peak is obtained also when one considers
just the single ordered sequence $\sigma^o$,
where each number can independently
be part of an IS with probability $p(T)=e^{1/T}/(e^{1/T}+1)$. Thus,
$\sigma^o$ represents $n$ independent paramagnets in a field, where
the variance of the length is just the sum of the single-number
variances $p(T)(1-p(T))$
and therefore
$C(T)=p(T)/(1-p(T))/T^2$ is readily available \footnote{As another
check of our algorithm, we sampled ISs for also $\sigma^o$ and
obtained the same $C(T)$ curve.}. This
$C(T)$ exhibits also a peak at the same temperature
$T\approx 0.4$. Thus, from the energetic
point of view no phase transition is visible. This is maybe similar
to spin glasses, where the transition to the RSB phase is also not
visible when studying the specific heat \cite{fischer1991}.

\begin{figure}
\includegraphics[angle=90,width=0.95\columnwidth]{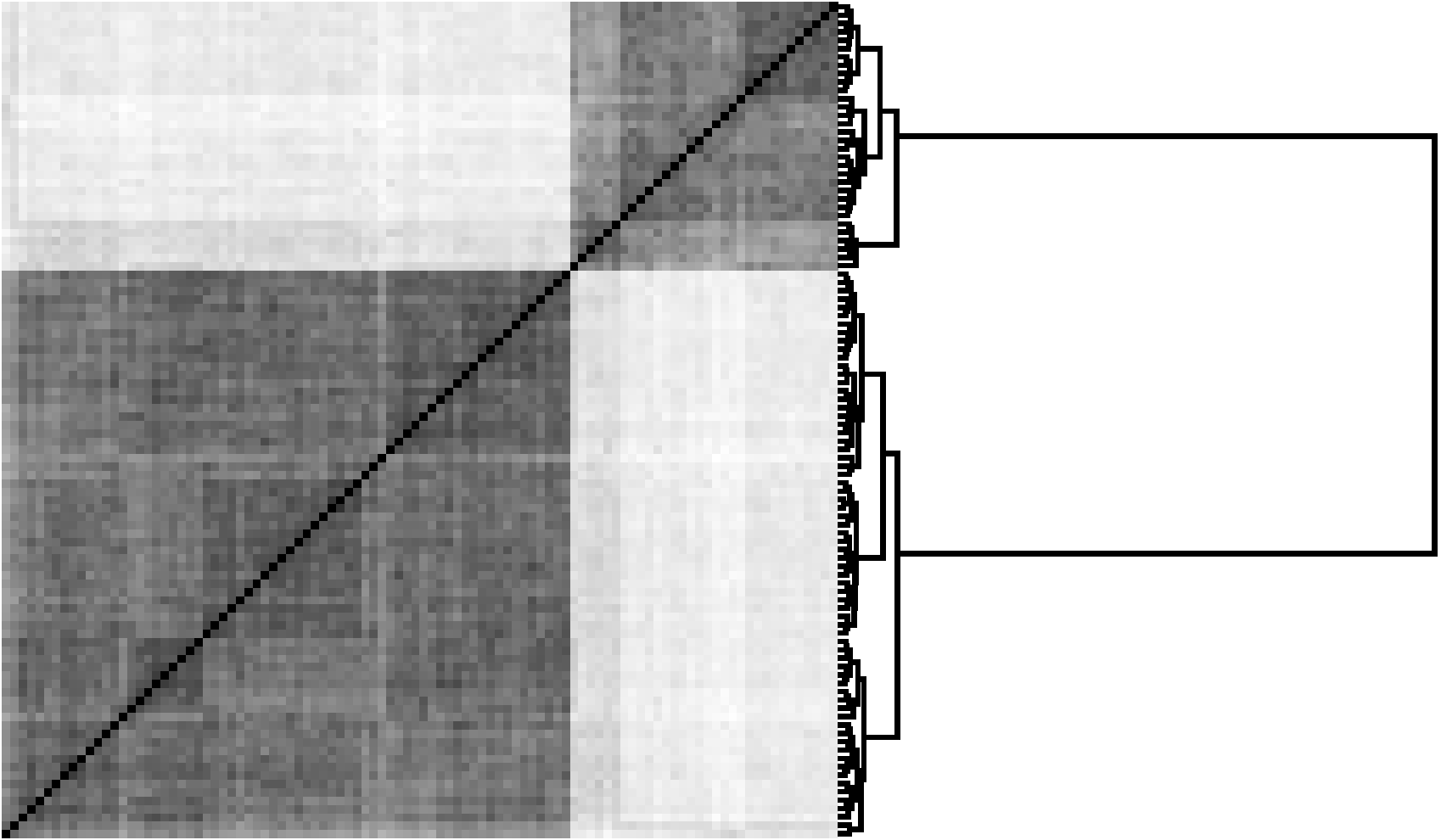}
\caption{\label{fig:dendrogram_perm} Clustered overlap matrix
  and dendrogram obtained by
  clustering  200 equilibrium IS sampled at
  $T=0.2$ for one random permutation with length $n=8192$.}
\end{figure}

The configuration space structure was further analyzed by applying the
\emph{agglomerative clustering approach} of Ward
\footnote{
The clustering approach \cite{ward1963,jain1988} operates on a set
of $M$ sampled configurations by initializing a set of $M$ clusters
each containing one configuration. One maintains pairwise distances
between all clusters, which are initially the distances $d=1-q$ between
the configurations, obtained from the overlaps $q$.
Then iteratively two clusters exhibiting the
currently shortest distance between them are selected and merged
to one single cluster, thereby reducing the cluster number by one.
For this new merged cluster, an updated distance to all other still
existing clusters have to be obtained. Here the update is done
with the approach of Ward \cite{ward1963}, which has
been used previously for the analysis of
disordered systems \cite{vccluster2004,1dchain_ultra2009,sat_cluster2010},
for more details see there.
The merging process is iterated until only one cluster is left.}.
The hierarchical structure obtained by the clustering
can be visualized by a tree, usually called
\emph{dendrogram}, where each branching corresponds to a subspace
of configurations,
see Fig.~\ref{fig:dendrogram_perm}.
The sequence of configurations as located in the leafs
defines a partial order.
This order can be used to display the matrix of the pair-wise
overlaps or distances  where the order of the rows and columns
is exactly given by the leaf order. The resulting matrix for
200 samples IS ($T=0.2$) of one random  permutation of length $n=8192$
is displayed in Fig.~\ref{fig:dendrogram_perm}. One observes
a hierarchical structure given by two major clusters,
visible by dark squares, i.e., similar configurations,
which are subdivided into sub clusters, with
relatively smaller similarities  on the off diagonals, respectively.

\begin{figure}
\includegraphics[width=0.99\columnwidth]{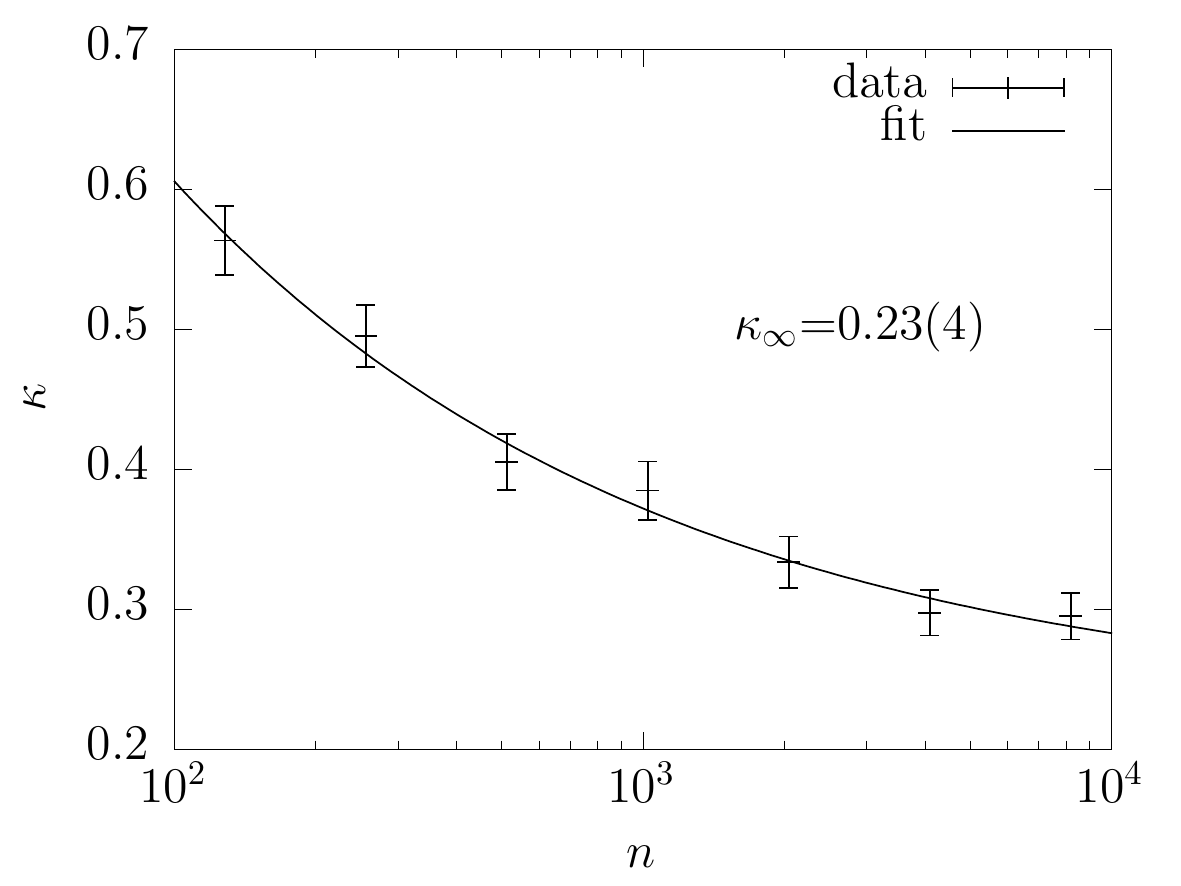}
\caption{\label{fig:kappa_N_perm} Average distance-dendrogram
correlation $\kappa$ as function of sequence length $n$,
for random permutations at $T=0.2$. The line shows a result when fitting
to a power law plus constant, see text.}
\end{figure}

The extend of the hierarchical structure can be made
quantitative by calculating the
\emph{cophenetic correlation}
\begin{equation}
{\kappa}\equiv [ d\cdot d_c]_P - [ d ] [
d_c ]_P\,,
\end{equation}
where $d=1-q$ is the distance corresponding to overlap value $q$.
The \emph{cophenetic distance}  $d_c$ between two states is
measured on the dendrogram as the distance
of the two largest clusters that contain only one of the states, respectively.
$[\ldots]_P$ denotes the combined average over the sampled IS and the
disorder ensemble.
Thus, this $\kappa$ measures the
correlation between the original distance $d$ of two states and the
distance $d_c$ imposed by the clustering, i.e., the degree of hierarchical
structure. In Fig.~\ref{fig:kappa_N_perm} $\kappa$ is shown as function
of $n$. By fitting a power law $\kappa(n)=\kappa_{\infty}+an^{-b}$ we
obtained $\kappa_{\infty}=0.23(4)$. This means, the IS landscape of
permutations exhibits also in the thermodynamic limit
a nested hierarchical structure, like it has been found for problems
exhibiting RSB as
mean-field spin glasses or some hard combinatorial optimization problems
\cite{vccluster2004}.

{\it (Summary and Discussion)}
The original problem of Ulam is to find the longest increasing subsequence for
random permutations. With so-far known algorithms it was possible to generate
one LIS, but in a statistically uncontrolled way. To study the
structure of configurations for LIS and IS for permutations,
we have introduced an algorithm which allows for exact and direct sampling of
increasing subsequences in polynomial time. For the uncorrelated and
most natural  ensemble of
permutations, we study the annealed mean and second moment of the number of
IS, which agree with the analytical calculations very well. We study
the structure of the configuration space and find a broad distribution of
overlaps and a hierarchical structure. By extrapolating
to infinitely long sequences, we show that this persists in the thermodynamic
limit, thus the model exhibits replica symmetry breaking. This model
is, to our knowledge, the first one which at the same time
exhibits RSB and for which
a polynomial exact sampling algorithm is available. This is in
strong contrast
to known complex but computationally hard problems like
mean-field spin glasses
and NP-hard optimization problems. Thus, Ulam's problem provides
an ideal test bed to study other phenomena of interest in the field
of complex disordered systems. In particular one can address
the non-equilibrium
behavior, the scaling of excitations, the coupling of replicas, or
an extended model obtained by the introduction of additional quenched
disorder, obtained by assigning individual local lengths for the
numbers. It could also be of interest to consider ensembles with
correlation or structure, in the spirit of a recent work
on directed polymers in random media \cite{directed_rsb2022}.
Furthermore, this study might motivate or help
to identify other models
with complex RSB behavior which can also be treated by polynomial algorithms.

\begin{acknowledgments}
  The simulations were performed at the
  the HPC cluster CARL, located at the University of Oldenburg
  (Germany) and
    funded by the DFG through its Major Research Instrumentation Program
    (INST 184/157-1 FUGG) and the Ministry of
    Science and Culture (MWK) of the
    Lower Saxony State.
\end{acknowledgments}

\bibliography{is_refs}

\end{document}